\documentclass[12pt,aps,prb,preprint]{revtex4}   

\usepackage{amsmath}    
\usepackage{graphicx}   

\begin{document}


\preprint{
    \vbox{
        \rightline{FSU-HEP 092209}\break
    }
}
\title{Varying-G Cosmology with Type Ia Supernovae}

\author{Rutger Dungan and Harrison B. Prosper}

\affiliation{Florida State University, Department of Physics, Tallahassee, FL 32306}

\begin{abstract}
The observation that Type Ia supernovae (SNe Ia) are fainter than expected given their red shifts
has led to the conclusion that the expansion of the universe is accelerating. The
widely accepted hypothesis is that
this acceleration is caused by a cosmological constant or, more generally, some dark energy field
that pervades the universe.  
In this paper, we explore what, \emph{on their own}, the supernovae data tell us about
this hypothesis. We do so by
answering the following question:
can these data be explained with a model in which the strength of gravity varies on a cosmic
timescale? We conclude that they can. Consequently, the supernovae data alone are insufficient to
distinguish between a model with a cosmological constant and one in which $G$ varies.
However,  the varying-G models prove not to be viable when other data are taken into account. This topic is an ideal one for investigation by 
an undergraduate physics major because the entire
chain of reasoning from models to data analysis is well within the
mathematical and conceptual sophistication of a motivated undergraduate.
\end{abstract}

\maketitle

\section{Introduction}

Physical cosmology~\cite{Weinberg, Hartle} is the branch of astronomy concerned with  the large scale structure and evolution of the universe. Cosmology is unique in that two apparently disjoint disciplines---the physics of the very small and the physics of the very large---are both needed to achieve our current understanding of the universe. It is remarkable, for example, that quantum fluctuations that occurred on microscopic scales in the very early universe may have left an imprint on the largest structures in the universe. The observation that the universe contains matter---with only trace amounts of antimatter, rather than matter and antimatter in equal amounts, may find its explanation, one day, using earthbound particle accelerators. Dark matter, for which there is much compelling evidence,~\cite{DarkMatter} may yet turn out to 
comprise weakly interacting particles that may be accessible in laboratories. The 
relatively recent synergy between
the theories of the very small and the very large is a thrilling achievement. However, there
is a cloud on the horizon called dark energy.~\cite{DarkEnergy, DE2}

A big surprise in cosmology came in 1998 when the High-Z Team~\cite{HiZ}
and the Supernova Cosmology Project~\cite{SCP} 
 both  observed, independently, that Type Ia supernovae  
were fainter than expected. After careful consideration of alternative explanations, 
both teams of researchers interpreted their observations as evidence that the SNe Ia  are further away than expected, given their red shifts and assuming a decelerating
universal expansion.  If the SNe Ia are further away than expected, then the average
expansion rate of the universe since the big bang must be higher than previously thought. 
Both teams, in fact, went further: they concluded that the
expansion of the universe is accelerating. Today, the broadly accepted 
hypothesis is that this acceleration is driven by a form of
energy called dark energy that pervades the universe. In the simplest model, dark energy is
identified with the cosmological constant $\Lambda$ that appears in the
general form of Einstein's theory of gravity, general relativity. In more complicated
models,~\cite{DE2} dark energy is modeled as a dynamical, evolving, field.

Cosmologists have created a 
compelling and coherent cosmology
based on the Friedmann equation
\begin{equation}
\left(\frac{\dot{a}}{a}\right)^2 = \frac{8 \pi G}{3} \rho  - \frac{Kc^2}{a^2} + \frac{\Lambda c^2}{3} ,
\label{eq:feq}
\end{equation} 
and the associated Friedmann-Lema\^{i}tre-Robertson-Walker (FLRW) metric~\cite{Weinberg} , 
\begin{equation}
ds^2 = c^2 dt^2 - a^2(t) \left[ \frac{dr^2}{1- K r^2} + r^2 (d\theta^2 + \sin^2\theta \, d\phi^2)\right],
\label{eq:metric}
\end{equation} 
where $a(t)$ is the dimensionless scale factor---normalized so that $a(t_0)=1$ at the present
time $t = t_0$, $\dot{a} \equiv da/dt$, $G$ is the gravitational constant, $\rho$  is the density of \emph{all}
forms of energy~\footnote
{The units of $\rho$, denoted $[\rho]$, are in fact \emph{mass} per unit volume $M L^{-3}$,
where $M$ is the mass unit and $L$ the unit of length.
However, it is common to choose units so that $c =  1$, thereby erasing 
the distinction between mass and energy. For pedagogical clarity however 
we keep the symbol $c$ in all expressions. Note: $[\Lambda] = $[K] = $L^{-2}$. }
excluding the contribution from the cosmological constant, $\Lambda$, 
and $-\infty < K < \infty$
is the spatial curvature. The radial coordinate $r$ is defined so  that the 
proper area of a sphere, centered at any conveniently chosen
origin, is $A_0 = 4 \pi r^2$ at the present time. As usual, symbols with a subscript of zero denote quantities evaluated
at  $t = t_0$.  The comoving distance $\chi$ associated with the radial coordinate
$r$ is given by
\begin{eqnarray}
\chi  	& = &  \int_0^r \frac{dr^{\prime}}{\sqrt{1 - K r^{\prime 2}}}, \nonumber \\
				& = & \sin^{-1} (K^{1/2} r) / K^{1/2},
				\label{eq:d0}
\end{eqnarray}
while $d(t) = a(t) \chi$ is the proper distance at time $t$. By construction, the comoving and proper distances are numerically identical today.
The radial coordinate $r$,  comoving distance $\chi$, radius of curvature $K^{-1/2}$, and 
proper distance $d(t)$ are conventionally 
measured in Mega-parsecs (Mpc). Inverting Eq.~(\ref{eq:d0}), we obtain
\begin{eqnarray}
r  	& = &  \sin(K^{1/2} \chi) / K^{1/2}.
	\label{eq:r}
\end{eqnarray}
For a spatially flat universe, that is, one with $K = 0$, Eq.~(\ref{eq:r}) simplifies to $r = \chi$. 

The standard model of cosmology, 
with $K = 0$ and
$\Lambda > 0$, works remarkably well; however, 
current physics predicts~\cite{DarkEnergy} a value of the cosmological constant $\Lambda$ that 
exceeds the observed value by a factor of at least $10^{50}$! This difficulty
motivates the exploration of  alternative explanations, such as ones that invoke
time-varying ``constants".~\cite{Barrow} After all, we know of no compelling reasons
why the parameters that appear in our current theories of the physical universe should
be independent of space and time. From some perspectives, the puzzle
is why they should be constant at all.~\cite{Martins} 

Another motivation for exploring alternative explanations of the supernovae data is to
determine whether they alone are sufficient to distinguish between a
model with a cosmological constant and models without, such as the varying-G models we
consider in this paper. 

A third, rather different motivation, is the pedagogical value of such investigations. This
topic is ideally suited for directed individual study (DIS) by an undergraduate
physics major. It is exciting and lends itself
to open-ended exploration. The work reported below was undertaken by one of the authors (RD),
an undergraduate physics major, under the supervision of the other (HBP).
We share the view of many that, ideally, all undergraduates
should be afforded the opportunity to engage in authentic research. However, this is
not always easy: many exciting topics unfortunately require rather more material
than can be 
mastered in a reasonable
amount of time by a busy student. 
The advantage of cosmology is that it is
intrinsically interesting to many students and, provided  one chooses the topic carefully and one
is prepared to make appropriate \emph{conceptual} approximations, one can find
interesting cosmological studies that can be done using mathematics
and concepts that are readily accessible to an undergraduate student.  We fully
endorse the idea that, for such students, a ``mathematics first" approach 
followed by applications is less desirable than the  ``physics first" approach as advocated by Hartle for general relativity.~\cite{Hartle} The 
cosmological investigation described below was done in that spirit.

This paper explores two simple phenomenological models
of varying-G cosmology~\cite{Barrow} using the data  compiled by Kowalski \emph{et al.}~\cite{SNIA} on 307 supernovae. We assume a spatially
\emph{flat} ($K =  0$) universe (motivated in part by the expectations from inflation~\cite{Weinberg})
and we set $\Lambda = 0$. However, for completeness, we write all expressions 
in a form that is valid for arbitrary values of $K$ and $\Lambda$. 

We find  fits to the supernovae data that are competitive with the  simplest dark energy model. The fact that non-dark energy models can 
account for these data  is a reminder that the supernovae
data alone are insufficient to establish dark energy  as the preferred hypothesis.  That hypothesis
becomes compelling only when different data-sets are analyzed together. Likewise, any
varying-G model must  fit not only the supernovae data,  but must also be in accord with other data. However, 
in view of our expressed goal to provide an example of an authentic research project that can be conducted in its entirety by an undergraduate student, we restrict the scope to only one
other datum: the bounds on  $\dot{G}/G$ at our current epoch. We
find that our two varying-G models  fail the bounds on  $\dot{G}/G$, thereby
ruling out this form of variation in $G$.
An interesting aspect of the first
model is that the scale factor becomes infinite 
 in a finite amount of time. In such a model, all comes to an end in a catastrophic shredding
of everything, a doomsday scenario that has been dubbed the \emph{big rip}.~\cite{bigrip}

The paper is organized as follows. Section~\ref{sec:cosmology} describes those 
additional aspects of 
FLRW cosmology that are needed to understand the SNe Ia data. Section~\ref{sec:varyg}
motivates the varying-G Friedmann equation we have used to describe the evolution of the
scale factor, while Sec.~\ref{sec:models} introduces our varying-G models and summarizes
the results we obtained from them. The paper ends with a discussion and concluding remarks.

\section{Supernova Cosmology}
\label{sec:cosmology}

A key problem in observational cosmology is measuring distances to galaxies. To do that we need two things: 1) a standard candle and 2) an operational definition of distance. We consider
first the standard candle. 

A standard candle is a source whose absolute luminosity is known. 
Type Ia supernovae~\cite{TypeI} are 
currently the best ``standardizable" candles for very large distances. A Type Ia supernova is
believed to occur when a star in a binary system overflows
its Roche lobe (the region within which its matter is gravitationally bound) causing material
from it to accrete onto the companion white dwarf. The mass of the white dwarf gradually increases towards the Chandrasekhar limit (of about 1.4 solar masses),~\cite{whitedwarf}
triggering runaway nuclear burning within the star that releases more energy
in a matter of weeks than the sun will emit in ten billion years. In another 
class of models, an explosion is triggered by the merger of two low-mass
white dwarfs. In a third class of models, a carbon-oxygen low-mass white dwarf 
explodes when the helium, accreted from a companion star, detonates. For a good review
of Type Ia supernovae models,  see Ref.~\cite{Hoeflich}.
By measuring specific
characteristics of the supernovae light curves (graphs of brightness as a function of time), it is possible to make empirically-derived corrections for the observed variations in SNe Ia
brightness and thereby create well-calibrated standard candles.~\cite{candles} 

\subsection{Luminosity Distance}
The proper distance between two points in space is a well-defined concept, but, unfortunately, it cannot be measured in practice. Instead, astronomers use
a definition of distance based on the flux of energy received on earth from
the luminous object, that is, on
the energy received per unit area per unit time,
\begin{equation}
f=\frac{L}{4\pi r^2} ,
\end{equation}
where $L = dE/dt$ is the object's luminosity, that is, its rate of total energy emission, 
and $A_0 = 4 \pi r^2$ is the proper area, at $t = t_0$,
of the sphere centered at the location once occupied by the supernova.  This formula
for the flux is valid for a static universe and for a source that emits energy isotropically.
In an expanding universe, however, the luminosity $L$ crossing this sphere is diminished by the factor $(1+z)^2$. One
factor of $1 + z$ arises
from the reduction in energy of each photon received on earth
relative to the energy it had at emission, yielding $dE \rightarrow dE / (1+z)$.   By definition, the red shift  $z \equiv (\lambda_r - \lambda_e)/\lambda_e$ where $\lambda_e$ and $\lambda_r$
are the emitted and received wavelengths, respectively. The second factor of $1 + z$ is due to
the reduction in the rate of arrival of photons at the earth, which yields 
$1/dt \rightarrow (1/dt)/(1+z)$.
The corrected expression for the flux is then 
\begin{equation}
 f = \frac{L}{4 \pi [(1+z)r]^2} \equiv \frac{L}{4 \pi d_L^2},
\end{equation}
where $d_L \equiv (1+z) r$ is called the luminosity distance. For arbitrary values of the 
curvature $K$, the radial coordinate $r$ is
related to the comoving distance $\chi$ via Eq.~(\ref{eq:r}), which, as noted
above, reduces to $r =  \chi$ when $K = 0$.

\subsection{Distance Modulus}
\begin{figure}[htbp]
\begin{center}
\includegraphics[scale=0.65] {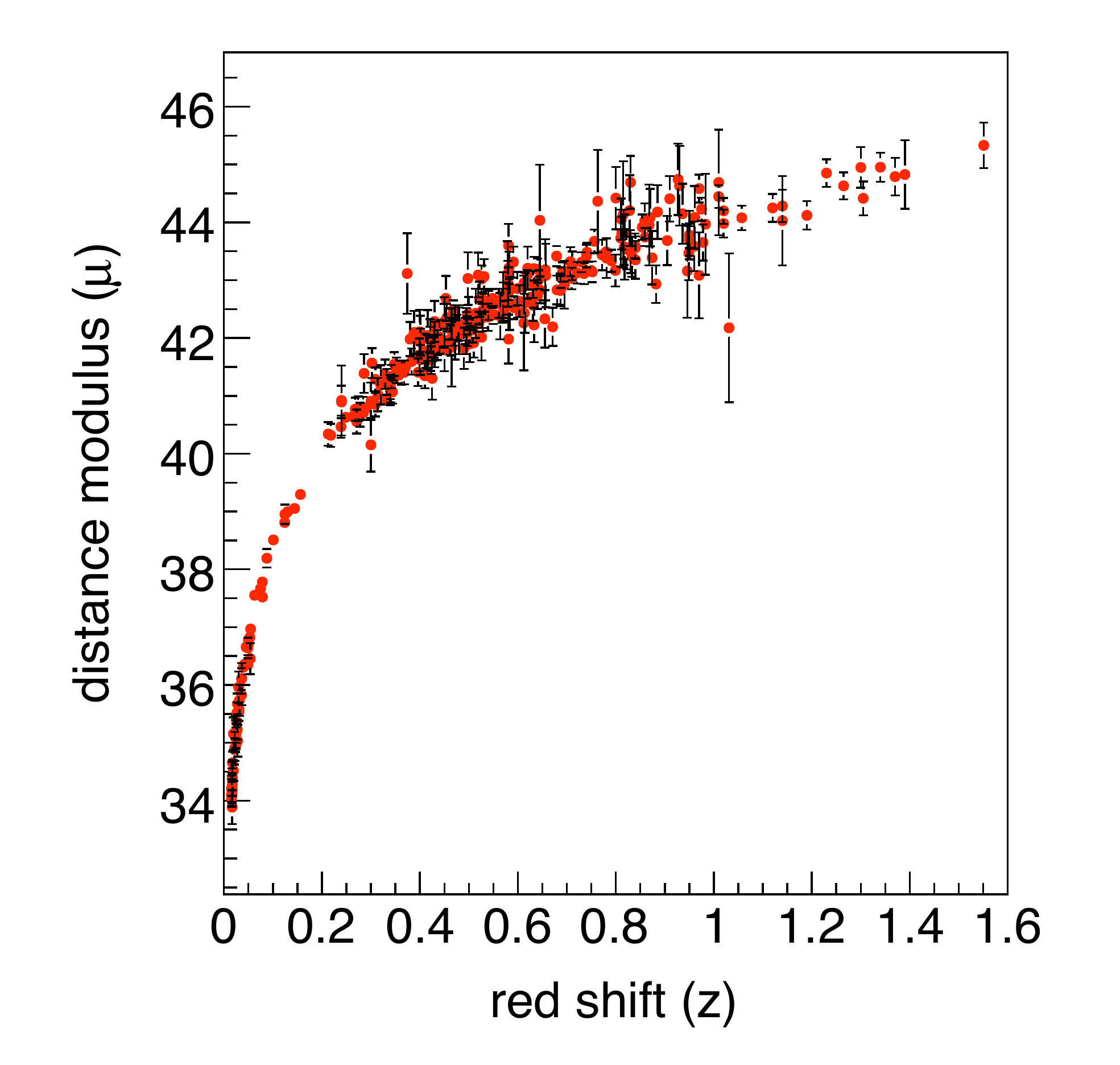}
\caption{Measurements of $(\mu, z)$ for 307 Type Ia supernovae, compiled 
by Kowalski \emph{et al.}}
\label{fig:sndata}
\end{center}
\end{figure}

Astronomers measure energy fluxes. But, by convention, fluxes are converted  into magnitudes 
$m$ defined by
$f = q 10^{-2m/5} = L/(4\pi d_L^2)$, where $q$ is the flux from objects of magnitude zero, with the luminosity distance  $d_L$ measured in Mega-parsecs (Mpc), and absolute magnitudes $M$ defined by
 $f_M = q 10^{-2M/5} = L/(4\pi d_M^2)$ where $d_M = 10^{-5}$ Mpc. Next they take the
 logarithm to base 10 of the ratio $f_M/f = 10^{0.4(m-M)} = (d_L/10^{-5})^2$ to arrive
 at
\begin{eqnarray}
 m - M & = & 5 \log_{10} [d_L /10^{-5}], \nonumber \\
            & = & 5 \log_{10} [(1+z) r(z)/10^{-5}].
\label{eq:modulus}
 \end{eqnarray} 
 Note that in the ratio the constant $q$ cancels. The difference $\mu \equiv m - M$ between the apparent magnitude $m$ of a source and its absolute magnitude
 $M$ is called the distance modulus. The analysis of a supernova light curve results
 ultimately in two measured quantities: $\mu$  and $z$. The data~\cite{SNIA} used in our study 
 are plotted in Fig.~\ref{fig:sndata}~\footnote{We have described the bolometric magnitude,
 that is, the magnitude of a star assuming we are able to be measure the flux
 across all wavelengths. In practice, however, fluxes are measured in wavelength bands defined by standard filters, such as the $B$-band filters.~\cite{UVB}
 An observed supernova spectrum is red shifted with respect to the spectrum in the rest
 frame of the supernova. Therefore, in general, a 
 filter will transmit a flux that differs from
 that which would have been measured were it possible to measure the flux in the supernova's rest frame. Astronomers use
 corrections called K corrections to
 map the measured flux back to its value in the 
 object's rest frame. Given
 a model of the object's spectrum, it is then possible, in principle, to infer the bolometric flux
 and hence the bolometric magnitude of the object. The distance moduli data compiled
 by Kowalski {\em et al.} are derived from $B$-band magnitude data.}. 
 The cosmology is all
 contained in the dependence of the radial distance $r$, or, equivalently, the comoving
 distance $\chi$, on the red shift
 $z$. The red shift, in turn, is related to the dimensionless scale factor, $a(t)$, as follows
 \begin{equation}
a = 1/(1+z).
\end{equation}

Given a functional relationship between the comoving distance $\chi$ and red shift $z$, the distance
modulus function in Eq.~(\ref{eq:modulus}) can be fitted to the data in Fig.~\ref{fig:sndata} to
extract the parameters of the cosmological model. A formula for the
comoving distance $\chi$ can be deduced from the FLRW metric, Eq.~(\ref{eq:metric}), by
noting that light in vacuum travels on null worldlines (for which $ds = 0$).
Therefore,
a light ray from a supernova at red shift $z$ satisfies the relationship
$c dt = a(t) d\chi$. Hence,  
\begin{eqnarray}
\chi(z) &  = & c \int_{(1+z)^{-1}}^1 \frac{da}{a \dot{a}}, 
\end{eqnarray}
noting that the light ray was emitted when the scale factor was $a = 1/(1+z)$ and received when
it assumes the value unity, today.

\section{A Varying-G Friedmann Equation}
\label{sec:varyg}

Our first assumption, which we have alluded to, is that the 
universe has zero spatial curvature.
Our second assumption is that the Friedmann equation, Eq.~(\ref{eq:feq}), for a $K = \Lambda = 0$
universe remains valid  
when $G$ is allowed to vary with time. By valid we mean that the equation is a good
approximation to some (unknown) exact equation describing the
evolution of the scale factor in a universe in which $G$ varies. This is an example of a
conceptual approximation that renders the problem tractable for an undergraduate
student. 
If we wish to remain strictly within the framework of general relativity,
we should be cautious about replacing Eq.~(\ref{eq:feq}) with one in which
$G$ is a function of time
because the Friedmann equation is derived from 
Einstein's equations 
\begin{equation}
G_{\mu\nu} +  \Lambda g_{\mu\nu} = \frac{8 \pi G}{c^4} T_{\mu\nu}, 
\end{equation}
which do not permit variations~\cite{Barrow} in $G$. The tensors
$G_{\mu\nu}$ and $T_{\mu\nu}$ are the components of the Einstein and
energy-momentum tensors, respectively, and $g$ is the metric tensor. 
In order to allow for a possible variation of
$G$, theories more general than Einstein's are needed, such as the scalar-tensor
theories~\cite{Barrow, Berro} in which gravity is assumed to couple to
a scalar field $\phi$, which---for weak constant coupling---yields the relationship $G \propto \phi^{-1}$.
This, in turn, yields a modified Friedmann equation with a time-dependent $G$ and 
additional terms
 of order $\dot{G}/G$. If the latter terms are small enough, we obtain a Friedmann equation
 identical in form to the standard one, but with a time-dependent $G$. 

Writing
$G(t) = G_0 f(a)$, where $G_0$ is the current value of $G$ and $f(a)$ describes the 
assumed dependence of $G$ on the scale factor $a(t)$ and therefore cosmic time $t$, 
and using the definitions 
\begin{eqnarray}
	\rho_{c0}		& \equiv & 3 H_0^2 / 8 \pi G_0, 				\nonumber \\
	\rho_{\Lambda}& \equiv & \Lambda c^2/8\pi G_0, 				\nonumber \\
	\Omega_M(a) 	& \equiv & \rho(a) / \rho_{c0},				\nonumber \\
	\Omega_{\Lambda} 	& \equiv & \rho_{\Lambda} / \rho_{c0}, 	\nonumber \\
	\Omega_0	& \equiv & \Omega_M(1) + \Omega_{\Lambda},
	\label{eq:defs}
\end{eqnarray}
where $\rho_{c0}$ 
is the critical density now, $\Omega_M(a)$ is the matter density parameter, 
and $H_0$ is the Hubble
constant---that is, the value of the Hubble parameter $H(t) \equiv \dot{a}/a$ today,
we may write the modified Friedmann equation as
\begin{equation}
\left(\frac{\dot{a}}{a}\right)^2  =  H_0^2 \left[ f(a) \Omega_M(a) + (1 - \Omega_0) a^{-2} + \Omega_{\Lambda}\right ],
                 \label{eq:friedmann}
\end{equation}
noting that   
\begin{equation}
-Kc^2 =  H_0^2 (1 - \Omega_0).
\label{eq:K}
\end{equation}
 With
these definitions, we can write the expressions for the comoving distance $\chi(z)$ and 
the universal time $t(a)$ as follows,
\begin{equation}
\chi(z)  =  \frac{c}{H_0}  \int_{(1+z)^{-1}}^1 \frac{da}{a^2 \sqrt{f(a) \Omega_M(a) + 
(1-\Omega_0) a^{-2} + \Omega_{\Lambda} }},
\label{eq:chi}
\end{equation}
and 
\begin{equation}
t(a) =  \frac{1}{H_0}  \int_0^a \frac{dx}{x \sqrt{f(x) \Omega_M(x) + 
(1-\Omega_0) x^{-2} + \Omega_{\Lambda} }}.
\label{eq:ta}
\end{equation}
The lifetime of the universe is given by $t_0 = t(1)$.

Our third assumption is that the \emph{total} mass-energy in the universe, whatever its
nature, scales in the same
way as matter; that is, we assume that $\Omega_{\Lambda} = 0$  and 
$\Omega_M(a) = \Omega_{M0} / a^3$, where $\Omega_{M0} = \Omega_M(1)$ 
denotes the value of the matter
density parameter today. Since we
also assume $K = 0$, Eqs.~(\ref{eq:defs}) and (\ref{eq:K}) show that, necessarily, 
$\Omega_0 = \Omega_{M0} = 1$. However, observations, interpreted
within the context of the standard cosmology~\cite{CMB} indicate that the
matter density parameter $\Omega_{M0}  \approx 0.3$. 
The difference between $\Omega_0 = 1$ and $\Omega_{M0}$ is presumed to be due to
the cosmological constant or dark energy. If we wished to be consistent with this
value of $\Omega_{M0}$, while keeping $\Lambda = 0$, we need to use a model with $K < 0$. 

One of our goals, however, is to ascertain whether the SNe data, 
\emph{on their own}, are sufficient to
conclude that the $\Lambda > 0$ model is preferred. To do so, we need
merely exhibit another model that works as well. Here we consider varying-G models
with $K = \Lambda = 0$ and therefore $\Omega_{M0} = 1$. Alternatively, one
could consider $\Lambda = 0$, $K \neq 0$, models. It should be noted, however, that the curvature term $(1- \Omega_0) a^{-2}$ cannot accelerate the expansion. In a universe dominated by  curvature, the Friedmann equation is $\dot{a} = \mbox{constant}$, which implies zero
acceleration. To obtain acceleration, one needs a term that dilutes less rapidly than the
curvature term,
which is the case for a cosmological constant or for the varying-G models described below.

\section{Varying-G Models and Results}
\label{sec:models}
In principle, a model for the variation of $G$ should arise from some deep
theory.~\cite{Barrow} This, however, is far beyond the scope of this paper,
which is to give an example of an interesting cosmological study than can be executed 
in its entirety by
an undergraduate physics major, but that nonetheless yields interesting results. We proceed in a purely
phenomenological manner. Our basic premise is that the supernovae are further away 
than expected because gravity was weaker in the past and, consequently, the universe
decelerated less rapidly than would be the case were $G$ constant and equal to its current value, $G_0$.  

\subsection{Fits to Supernovae Data}
We studied several forms for the function $f(a)$ in 
$G(a) = G_0 f(a)$, but in this paper we report results for only two of them, each with a 
\emph{single} adjustable, dimensionless, parameter, $b$.
The first varying-G model we studied is defined by
\begin{equation}
f(a) = e^{b(a-1)}, \, \, \, \, \mbox{model 1}. \nonumber
\end{equation}
In this model, there is no limit to how strong gravity can become. Another model studied is
defined by
\begin {equation}
f(a)=2/(1+e^{-b(a-1)}),  \, \, \, \, \mbox{model 2}, \nonumber
\end{equation}
in which $G$ is limited to twice its current value in the distant future. 
We have normalized both models so that $G(a)$ assumes its current value, $G_0$, when $a = 1$. 
For $K = 0$ models, the distance modulus, Eq.~(\ref{eq:modulus}), may be
written as 
\begin{equation}
\mu(z, b, Q) = 5 \log_{10} [(1+z)  H_0 \, r(z)/c] + Q,
\label{eq:mu}
\end{equation}
where 
the offset $Q$ determines the vertical location of the modulus curve~\footnote{The absolute luminosity of a supernova cannot be determined
independently of the Hubble constant $H_0$. Consequently, in the fit of the modulus function to
the data, it is only the \emph{shape} of the function that contains useful information
about the cosmology. The offset $Q$ depends both on $H_0$ as well as on
the flux corrections.}. 
Note that $H_0 \, r(z)/c$ is dimensionless
and independent of the Hubble constant.

Evaluating Eqs.~(\ref{eq:chi}) and (\ref{eq:ta}) for model 1, with $K = 0$ (that is, $\Omega_0 = 1$) and 
$\Omega_{\Lambda} = 0$, we find
\begin{equation}
\chi(z) = r(z) =  \frac{c}{H_0}  e^{b/2} \sqrt{2\pi/b} [\mbox{erf}(\sqrt{b/2}) - \mbox{erf}(\sqrt{b(1+z)^{-1}/2}) ],
\end{equation}
and 
\begin{equation}
t(a) =  \frac{1}{H_0} [e^{b/2}( \sqrt{2\pi/b} \, \mbox{erf}(\sqrt{ab/2}) - 2\sqrt{a} \, e^{-ab/2}) ] / b.
\label{eq:lifetime}
\end{equation}
This model exhibits a striking feature: the scale
factor becomes infinite in a finite amount of time. We shall
return to this point below. For model 2, the integrals in 
Eqs.~(\ref{eq:chi}) and (\ref{eq:ta}) are evaluated
numerically using 
the mid-point rule.~\footnote{If the interval $[a,b]$ is divided into $N$ intervals
of width $h = (b-a)/N$, the mid-point rule is
$\int_a^b f(x) \, dx \approx h \sum_{i=1}^N f(a + (i-0.5) h)$.}
 
\begin{figure}[htbp]
\begin{center}
\includegraphics[ scale=0.65] {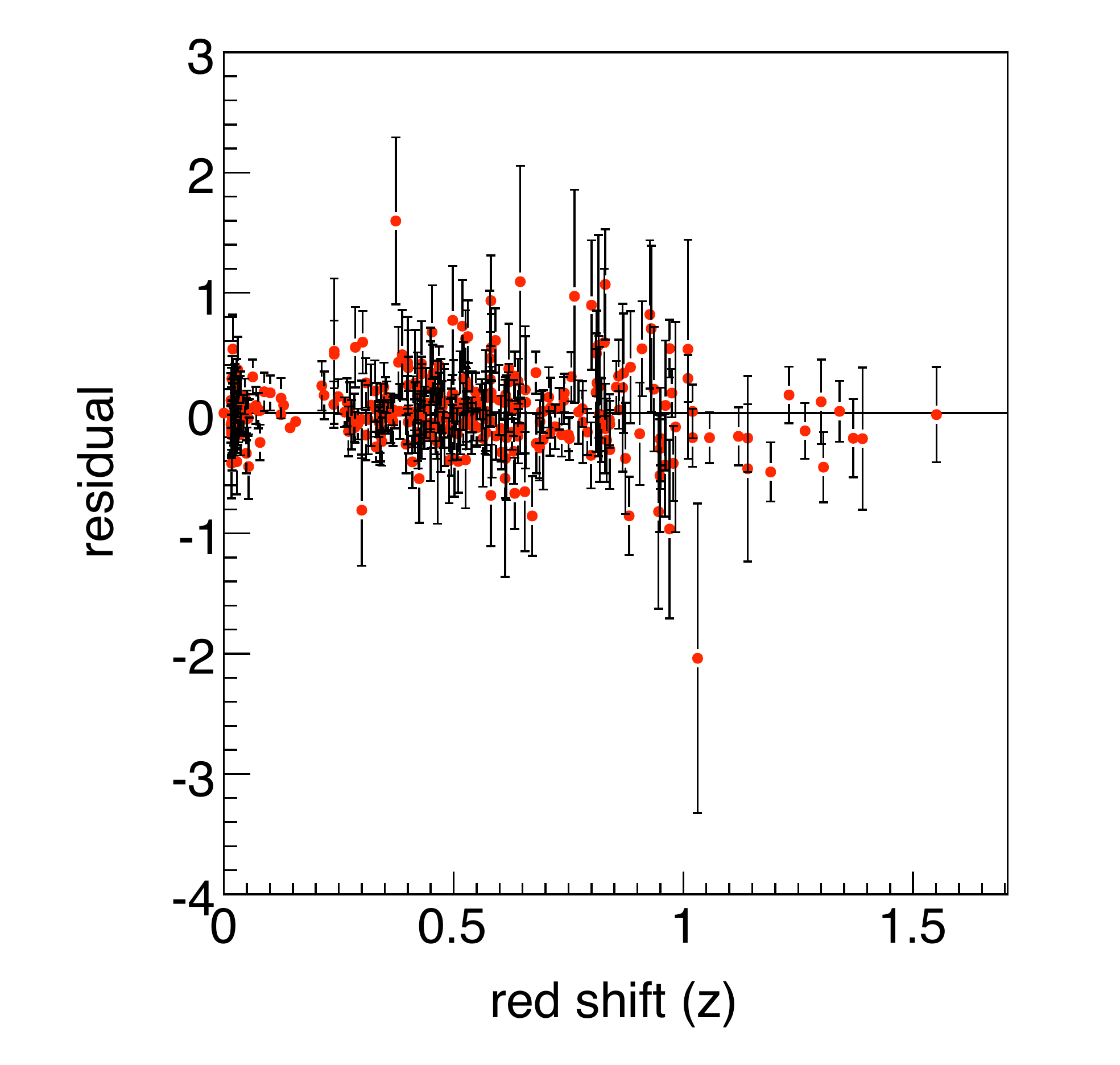}
\end{center}
\caption{Residuals, $\mu_n - \mu(z_n)$, of fit of model 1 to SNe Ia data ($\chi^2/ND = 313.0/305 = 1.03$).}
\label{fig:model1}
\end{figure}

\begin{figure}[htbp]
\begin{center}
\includegraphics[scale=0.65]{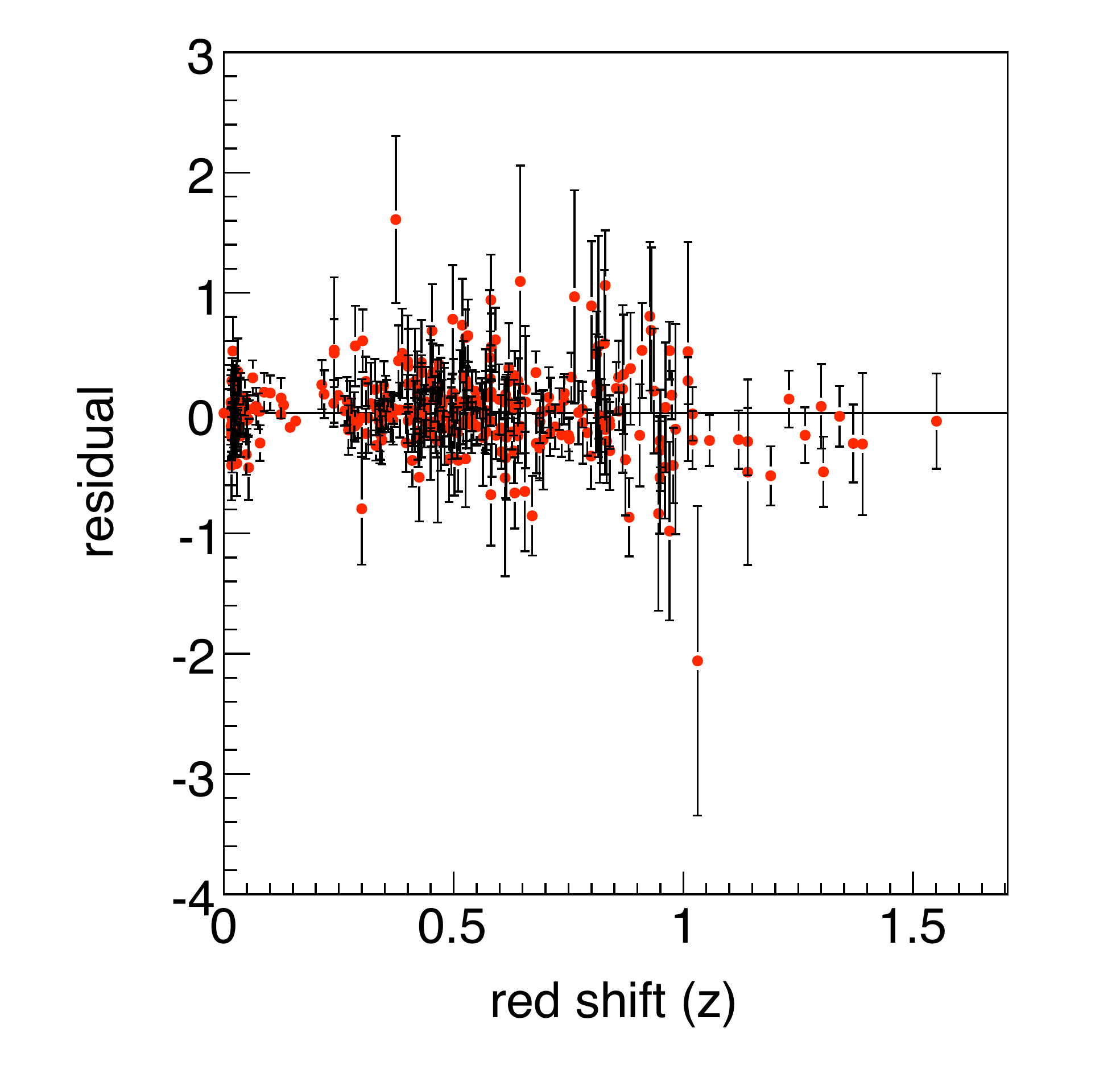}
\caption{Residuals, $\mu_n - \mu(z_n)$, of fit of model 2 to SNe Ia data ($\chi^2/ND = 316/305=1.04$).}
\label{fig:model2}
\end{center}
\end{figure}

We fit Eq.~(\ref{eq:mu}) to the SNe data in Fig.~\ref{fig:sndata}
by minimizing the function
\begin{equation}
\chi^2 = \sum_{n=1}^{307} [\mu_n - \mu(z_n, b, Q) ]^2 / \sigma_n^2,
\label{eq:chi2}
\end{equation}
with respect to the parameters $b$ and $Q$, where $1 \leq n \leq 307$ 
labels the $n^{\mbox{th}}$ supernova at red shift $z_n$ and
distance modulus $\mu_n$, measured with an uncertainty of $\pm \sigma_n$. The minimization of Eq.~(\ref{eq:chi2}) is
done using the program TMinuit, which is part of the  ROOT data analysis package from CERN.~\cite{ROOT} For model 1, we get the 
result shown
in Fig.~\ref{fig:model1}.
The fit gives the value $b = 2.09 \pm 0.08$,
from which we infer a lifetime of 
$t_0 = 15.1 \pm 0.3$ (70 km s$^{-1}$Mpc$^{-1}$/$H_0$) Gyr.~\footnote{
The lifetimes can be computed given values for the parameters $b$ and $H_0$. 
However, since we cannot extract a value of $H_0$ from the fits, we compute the lifetimes 
using the nominal value 70 km s$^{-1}$Mpc$^{-1}$ for the Hubble constant, 
but we write all lifetimes in terms of the parameter $H_0$ to make clear how
the numerical values will change if $H_0$ differs from the nominal value.}  
The fact that the $\chi^2$ per degree of freedom ($\chi^2/ND)$ is 1.03
suggests that the modulus uncertainties are estimated
correctly and that model 1 provides an excellent description of the 
data.~\footnote{If the reported modulus uncertainties 
are Gaussian distributed, one expects $\chi^2$ to be sampled from a probability density with mean $N - P \equiv ND$, where $N = 307$ is the number of data points and $P \approx 2$ is the number of adjustable 
parameters. Therefore, for a fit that neither over-fits nor under-fits, 
we expect $<\chi^2>/ND \approx 1$. The quantity $P$ would be exactly equal to 2 if the constraints
that define the parameter estimates were \emph{linear} in the parameters. }
A similarly good fit is found for model 2, as shown in Fig.~\ref{fig:model2}.
This fit yields $b = 3.27 \pm 0.11$, with a $\chi^2/ND$ = 316/305 = 1.04. 
We find $t_0 = 16.2 \pm 0.4$ (70 km s$^{-1}$Mpc$^{-1}$/$H_0$) Gyr. 
 For the simplest dark energy model, for which
 $f(a) = 1$ and $\Omega_M(a) = (1-\Omega_{\Lambda})/a^3$ with $\Omega_{\Lambda} > 0$,
 we find  $\Omega_{\Lambda} = 0.71 \pm .02$ and 
  $t_0 = 14.0 \pm 0.3$ (70 km s$^{-1}$Mpc$^{-1}$/$H_0$) Gyr,
 consistent
with the accepted results.~\cite{DarkEnergy}
The 
$\chi^2$ per degree of freedom of the fit is 
310/305 = 1.02.

Since there is no compelling statistical basis to reject any of these models, 
we conclude that the supernovae data alone are insufficient to distinguish between
them. However, these data when analyzed along with others~\cite{DarkEnergy}
are consistent with  a simple cosmology in which dark energy mimics a cosmological
constant with the value  $\Omega_{\Lambda} \approx 0.7$. The varying-G models should
likewise be analyzed along with other data to see if a consistent picture emerges. 
The fact that a $K = \Lambda = 0$ model requires $\Omega_{M0} = 1$, while the
preferred value from galaxy and galaxy cluster measurements is $\Omega_{M0} = 0.3$,
is already an indication of a difficulty. A systematic analysis of the relevant data, however, is
a large task beyond the scope of this paper. Instead, we 
illustrate the importance of including other data by comparing the predicted
fractional 
variation of $G$, $\dot{G}/G$ at the present epoch, with the available bounds. 

\subsection{Bounds on the Variation of $G$}
\begin{figure}[thb] 
\begin{center}  
\includegraphics[scale=0.65]{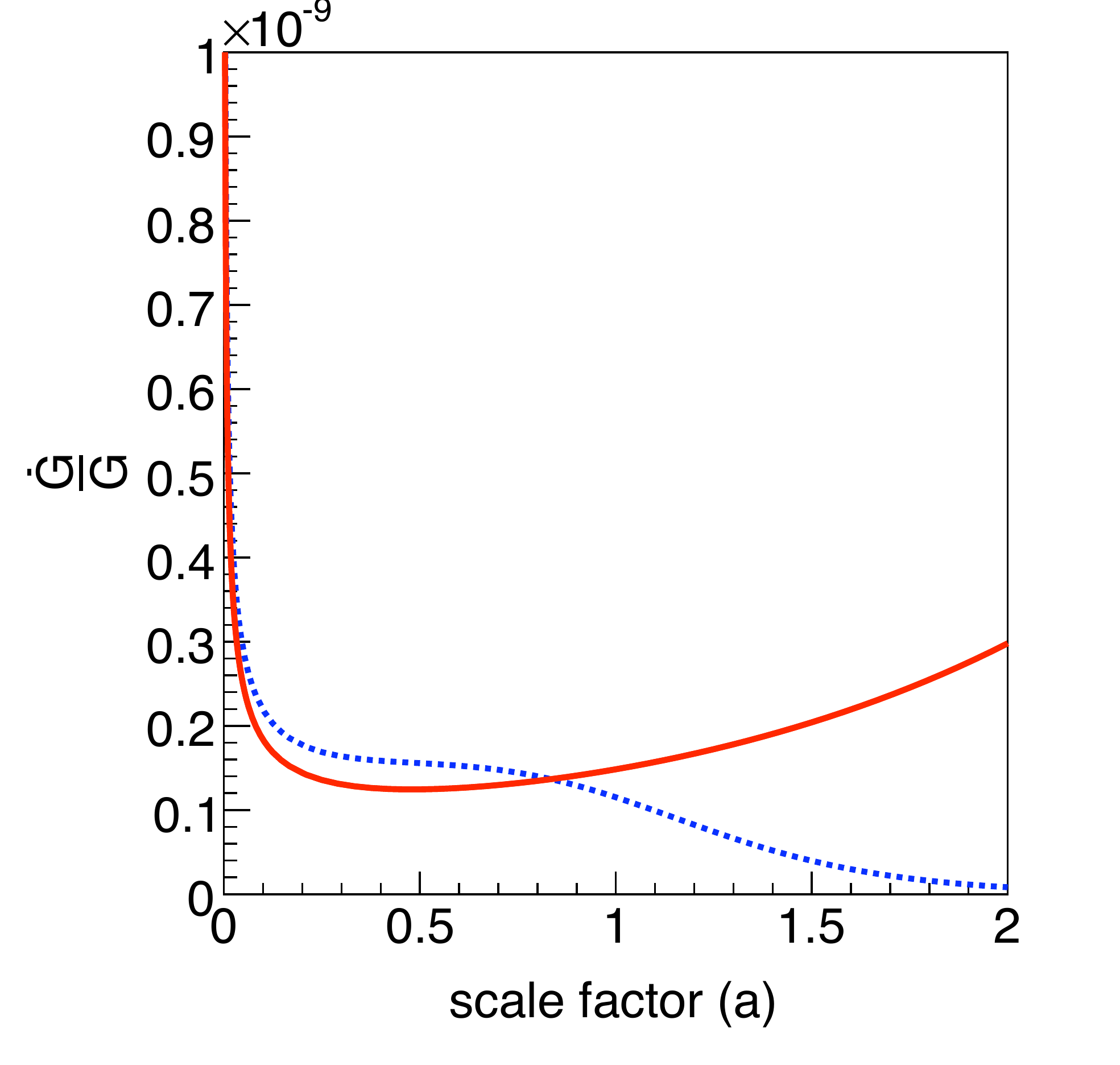} 
\caption{Fractional change in $G$ per year for model 1 (solid curve) and model 2 (dashed curve) as a function of the 
dimensionless scale factor
$a(t)$.}
\label{fig:GdotG}
\end{center}
\end{figure}

The possible variation of $G$ is usually characterized by the quantity $\dot{G}/G$,
which  in terms of the logarithmic derivative of the function $f(a)$ is given by
\begin{equation}
\frac{\dot{G}}{G}   =  \frac{1}{G} \frac{dG}{da} \dot{a}
                               = H_0 \frac{d\ln f}{da},
\end{equation}
where we have used the fact that $H_0 = \dot{a}/a = \dot{a}$ at the present epoch.
Figure~\ref{fig:GdotG} shows $\dot{G}/G$ as a function of the scale factor, for models 1 and 2.
We see that at $a=1$,   
$\dot{G}/G$ is equal to  $1.5 \times 10^{-10}$ y$^{ -1} $ and $1.15 \times 10^{-10}$ y$^{ -1} $,  respectively. Unfortunately, these values for $\dot{G}/G$ are one to three orders of magnitude
larger than the upper bounds that range from about $10^{-10}$ y$^{-1}$ to
$10^{-13}$ y$^{-1}$ depending on the method used to extract the bound.~\cite{GdotG} 
 
\section{Discussion and Conclusion}

Engaging undergraduate students in
research can be an effective way to keep them excited about science. Cosmology is particularly
well suited in this regard because it is possible to find topics that are both
manageable in scope and scientifically interesting. We have presented an
investigation of varying-G cosmological models that serve as examples
of interesting research problems that are well matched to the mathematical sophistication of an undergraduate. 

The two phenomenological models we presented, in which the strength of gravity
{\em increases} with cosmic time, provide excellent fits to the Type Ia supernovae data. We therefore conclude that the supernovae data alone cannot establish the dark energy
hypothesis unambiguously. Other data are needed to render this hypothesis plausible. However, both our varying-G models fail to satisfy the bounds on 
$\dot{G}/G$. Consequently, the particular variation of $G$ described by these models is
ruled out.  
In fact, one can make a stronger statement: \emph{all} varying-G
models that give rise to accelerated expansion, and that are
based on the FLRW metric and the Friedmann equation, are ruled out by these bounds.~\cite{Linder}
Consider, for example, matter-dominated models, for which the Friedmann
equation is $H^2 \sim G /a^3 \rightarrow G  \sim a \dot{a}^2$. This yields $\dot{G}/G \sim H + 2\ddot{a}/\dot{a}$, from which we conclude that $\dot{G}/G \gtrsim H$. But a value of
$\dot{G}/G$ of the order of $H_0$ is inconsistent with
the bounds on $\dot{G}/G$, which are \emph{less} than $H_0$ by one to three orders of magnitude. 

The inability to distinguish between models (model degeneracy) is inherent in the Friedmann
equation because the latter is sensitive only to the \emph{total} energy density of the universe
and is agnostic with respect to how the energy density arises. This may be seen
by writing the Friedmann equation, Eq.~(\ref{eq:feq}), as
\begin{eqnarray}
\left( \frac{\dot{a}}{a}\right)^2 & = & \frac{8\pi G}{3} (\rho_M + \rho_K + \rho_{\Lambda}),
\end{eqnarray}
where $\rho_M$, $\rho_K$, and $\rho_{\Lambda}$ are the density contributions from
matter, the curvature and the cosmological
constant, respectively. The Hubble parameter $\dot{a}/a$ is related to the \emph{sum} of
$\rho_M + \rho_K + \rho_{\Lambda}$ not the individual components;  
or, equivalently, to the sum 
$$\Omega \equiv f(a) \Omega_M(a) + (1-\Omega_0) a^{-2} + \Omega_{\Lambda}.$$
Therefore, it is 
possible to
entertain different interpretations of the total energy density. For example, \emph{any} model based
on the Friedmann equation
can be reinterpreted as one in which matter, perhaps of several different sorts,
is either \emph{created}, \emph{destroyed}, or both, as the universe evolves. Consider,
for example, the simple cosmological constant dark energy model, for which  $f(a) = 1$,
$\Omega_0 = 1$ and the total energy density 
is given by $\Omega = (1 - \Omega_{\Lambda})/a^3 + \Omega_{\Lambda}$.
This can be rewritten
as $\Omega = \Omega_M^{\prime}(a)/a^3$ with $ \Omega_M^{\prime}(a) = 1 - \Omega_{\Lambda} + \Omega_{\Lambda} a^3$. Since $a^{-3}$ is the dilution factor for matter, the function $\Omega_M^{\prime}(a)$ describes an increasing matter density in a comoving
volume, which can be interpreted as the creation of matter as the
universe expands! Alternatively, as done here, one can maintain the mass continuity
equation, in which case matter is neither created nor destroyed and
$\Omega = \Omega_{M0}/a^3$, but allow $G$ to vary  like 
$f(a) \sim 1 + a^3 (1 - \Omega_{M0})/\Omega_{M0}$. Because of the invariance of the
Friedmann equation with
respect to such changes in interpretation, it is necessary to impose constraints on the cosmological parameters to remove the model degeneracy. Such constraints can come from
other data, or other equations, or both. 

It seems odd, at first, that the strengthening of gravity with time leads not to the eventual
gravitational collapse of the universe, but rather to its accelerating  expansion. The reason
for this is that every form of energy contributes to the geometry of
spacetime. A model in which the
strength of gravity changes with time is equivalent to another model in which the energy density 
changes in a specific way. 
If the energy density dilutes more rapidly than $a^{-2}$, 
then the expansion will slow down. If the strength of gravity increases such that in the equivalent,
constant-G model, the energy dilutes more slowly than $a^{-2}$, the expansion will accelerate.
In our varying-G models, the effective energy density increases with time. 

For
model 1, the increasing strength of gravity leads to a startling prediction:
a catastrophic end to such a universe. 
This conclusion follows from the limit 
$a \rightarrow \infty$ of the lifetime expression, Eq.~(\ref{eq:lifetime}). We find that
\begin{equation}
 t(a \rightarrow \infty) = \frac{1}{H_0}  \exp(b/2)\sqrt{2\pi/b}/b.
\end{equation}
According to this model, the universe has a finite lifetime of about 33 Gyr and
will tear itself to pieces in  its final moments!
Such behavior has been dubbed the {\em big rip} and is a feature of cosmological
models containing phantom energy.~\cite{bigrip}  Within regions that are dominated by non-gravitational
forces, the effect of a cosmological constant does not change with time and consequently the accelerating
universal expansion will not disrupt already bound systems. By contrast, as the universe
ages the effect of phantom energy increases in {\em any} finite volume of space. Eventually, this precipitates an escalating cascade of destruction at ever smaller scales until
everything is torn asunder.  We can only hope that phantom energy is just that: a phantom!

\section*{Acknowledgements}
We thank Peter H\"{o}flich for an insightful discussion on the 
physics of Type Ia supernovae. This work was supported in part by
a grant from the U.S. Department of Energy.

\end{document}